\documentclass[twocolumn,pre,superscriptaddress,aps,footinbib]{revtex4}
\usepackage[usenames,dvipsnames]{color}
\usepackage{url,psfrag,graphicx}
\usepackage{dcolumn}
\usepackage{amsmath,amssymb}
\usepackage{bm}
\usepackage{hyperref}
\usepackage{float,epsfig,color}
\usepackage{times}
\usepackage[latin1]{inputenc}

\usepackage{graphicx}
\usepackage{amsmath}
\usepackage{amssymb}
\usepackage{dcolumn}

\begin{document}

\title{Anomalous ballistic scaling in the tensionless or inviscid Kardar-Parisi-Zhang equation}

\author{Enrique Rodr\'{\i}guez-Fern\'andez}
\email{enrodrig@math.uc3m.es}
\affiliation{Departamento de Matem\'aticas and Grupo Interdisciplinar de Sistemas Complejos (GISC)\\ Universidad Carlos III de Madrid, Avenida de la Universidad 30, 28911 Legan\'es, Spain}
\author{Silvia N.\ Santalla}
\email{silvia.santalla@uc3m.es}
\affiliation{Departamento de F\'{\i}sica and GISC, Universidad Carlos III de Madrid, Avenida de la Universidad 30, 28911 Legan\'es, Spain}
\author{Mario Castro}
\email{marioc@iit.comillas.edu}
\affiliation{Instituto de Investigaci\'on Tecnol\'ogica (IIT) and GISC, Universidad Pontificia Comillas, 28015 Madrid, Spain}
\author{Rodolfo Cuerno}
\email{cuerno@math.uc3m.es}
\affiliation{Departamento de Matem\'aticas and Grupo Interdisciplinar de Sistemas Complejos (GISC)\\ Universidad Carlos III de Madrid, Avenida de la Universidad 30, 28911 Legan\'es, Spain}

\begin{abstract}
The one-dimensional Kardar-Parisi-Zhang (KPZ) equation is becoming an overarching paradigm for the scaling of nonequilibrium, spatially extended, classical and quantum systems with strong correlations. Recent analytical solutions have uncovered a rich structure regarding its scaling exponents and fluctuation statistics. However, the zero surface tension or zero viscosity case eludes such analytical solutions and has remained ill-understood. Using numerical simulations, we elucidate a well-defined universality class for this case that differs from that of the viscous case, featuring intrinsically anomalous kinetic roughening, despite previous expectations for systems with local interactions and time-dependent noise and ballistic dynamics. The latter may be relevant to recent quantum spin chain experiments which measure KPZ and ballistic relaxation under different conditions. We identify the ensuing set of scaling exponents in previous discrete interface growth models related with isotropic percolation, and show it to describe the fluctuations of additional continuum systems related with the noisy Korteweg-de Vries equation. {Along this process, we additionally elucidate the universality class of the related inviscid stochastic Burgers equation.}
\end{abstract}

\maketitle

\section{Introduction}
\label{sec:intro}

The Kardar-Parisi-Zhang (KPZ) equation describing the evolution of a front with height $h(x,t)$ above position $x\in \mathbb{R}$ along a one-dimensional (1D) substrate at time $t$ reads \cite{Kardar86}
\begin{align}
    \partial_t h = \nu \partial^2_x &h + \frac{\lambda}{2} (\partial_x h)^2 + \eta(x,t) , \label{eq:kpz} \\
    \langle \eta(x,t) \eta(x',t') \rangle &= 2D \, \delta(x-x') \delta(t-t') ,
    \nonumber
\end{align}
where $\nu, D>0$, and $\lambda$ are parameters and $\eta(x,t)$ is zero-average, Gaussian white noise. Within the physical picture of a growing interface that led to the seminal proposal of Eq.\ \eqref{eq:kpz} in Ref.\ \cite{Kardar86}, the nonlinear term approximates growth of the front at a constant rate along the local normal direction, the noise implements the inherent stochasticity of microscopic growth events in time and space, and the diffusive linear term represents smoothening mechanisms, like surface tension, which reduce the local height differences \cite{Barabasi95,Krug97}. Alternatively, the space derivative of Eq.\ \eqref{eq:kpz} yields for $u=\partial_x h$ the Burgers equation for the velocity of a stochastically driven incompressible fluid, where $\nu$ measures viscosity \cite{Kardar86,Rodriguez-Fernandez19,Rodriguez-Fernandez20}.

The KPZ equation is a nonequilibrium system that, irrespective of parameter values, is well known to display generic scale invariance \cite{Grinstein95,Tauber14}, termed kinetic roughening in the surface growth picture \cite{Barabasi95,Krug97}. The universality class it represents is very recently proving ubiquitous for low-dimensional systems with strong fluctuations, from non-quantum contexts like turbulent liquid crystals \cite{Takeuchi11}, stochastic hydrodynamics \cite{Mendl13}, colloidal aggregation \cite{Yunker13}, reaction-diffusion systems \cite{Nesic14}, random geometry \cite{Santalla15,Santalla17}, active matter \cite{Chen16}, or thin films \cite{Orrillo17}, to the quantum realm, as for superfluidity \cite{Altman15}, entanglement \cite{Nahum17}, electronic fluids \cite{Protopopov21}, or integrable and non-integrable quantum spin chains \cite{Gopalakrishnan19,Ljubotina19,DeNardis21}. E.g., super-diffusive transport of quantum spin excitations has been recently measured experimentally \cite{Wei21}, finding values for the dynamic exponent $z<2$ ---that quantifies the increase of the correlation length, $\xi(t) \sim t^{1/z}$ \cite{Barabasi95,Krug97}--- consistent with the 1D KPZ behavior ($z_{\rm KPZ}=3/2$), or else with ballistic transport, $z_{\rm b}=1$.

An important role in the recent identification of 1D KPZ scaling for so many different physical contexts has been played by the analytical solutions of Eq.\ \eqref{eq:kpz} \cite{Sasamoto10,Amir11,Calabrese11} and of discrete models in the same universality class, see reviews e.g.\ in Refs.\ \cite{Kriechebauer10,Corwin12,Halpin-Healy15,Takeuchi18}. Beyond exact values for the scaling exponents, such analytical solutions include the covariance and probability distribution function (PDF) of the height fluctuations, and their dependence on global constraints like constant vs time-dependent system size \cite{Kriechebauer10,Corwin12,Halpin-Healy15,Takeuchi18}.
For the KPZ equation itself, a key step towards the exact solution is the Cole-Hopf transformation $H(x,t)=\exp[\lambda h(x,t)/(2\nu)]$ \cite{Kardar86}, which converts Eq.\ \eqref{eq:kpz} into the stochastic heat equation for $H(x,t)$ \cite{Kriechebauer10,Corwin12,Halpin-Healy15,Takeuchi18}. However, this mapping is unavailable in the tensionless case when $\nu=0$, namely, for the equation
\begin{equation}\label{eq:invkpz}
    \partial_t h = \frac{\lambda}{2}(\partial_x h)^2 + \eta(x,t) .
\end{equation}
Equation \eqref{eq:invkpz} has been considered as a model of conserved relaxation \cite{Golubovic91}. Being marginally unstable to perturbations of a flat solution \cite{Golubovic91,Cuerno95}, it is arduous to integrate numerically, leading to questions on its well-posedness \cite{Tabei04,Bahraminasab04}. In this {paper}, a suitable numerical algorithm allows us to uncover a well-defined universality class for Eq.\ \eqref{eq:invkpz}. The scaling exponents differ from those of the 1D KPZ equation with $\nu\neq 0$ and correspond to ballistic relaxation with $z=z_{\rm b}$. Moreover, the usual dynamic scaling Ansatz satisfied in the $\nu\neq0$ KPZ case does not hold. Rather, intrinsic anomalous scaling \cite{Schroeder93,DasSarma94,Lopez97,Ramasco00,Cuerno04} occurs, whereby the scaling exponents differ for local and global fluctuations. Thus, Eq.\ \eqref{eq:invkpz} yields a counterexample for an almost two-decades old conjecture \cite{Lopez05} that intrinsic anomalous scaling cannot be asymptotic for continuous models with local interactions and spatiotemporal noise. Notably, the universality class thus uncovered for Eq.\ \eqref{eq:invkpz} has been reported earlier for discrete growth models related with isotropic percolation \cite{Asikainen02,Asikainen02b}. Furthermore, we obtain it here for yet another paradigmatic continuous system related with the stochastic Korteweg-de Vries (KdV) equation with time-dependent noise. {Along this process, we additionally elucidate the universality class of the related inviscid stochastic Burgers equation.}

{This paper is organized as follows. In section \ref{sec:scaling}, the kinetic roughening of both, the tensionless KPZ and the inviscid Burgers equations, are described in detail. An indirect integration of the tensionless KPZ equation is also assessed in Appendix \ref{app:1}, as a cross-check. The statistics of the fluctuations for both equations are studied in section \ref{sec:fluctuation}. A final discussion of the universality classes elucidated by our present work is addressed in section \ref{sec:universality}. This includes additional numerical simulations of different versions of the stochastic KdV equation, which prove them as further conspicuous members of these universality classes. Finally, our conclusions and an outlook are presented in section \ref{sec:conclusions}.}

\section{Kinetic roughening: scaling Ansatz and exponents }\label{sec:scaling}

\subsection{Tensionless KPZ equation: direct simulations}
\label{sec:direct}

\begin{figure}[!t]
\begin{center}
\includegraphics[width=1.00\columnwidth]{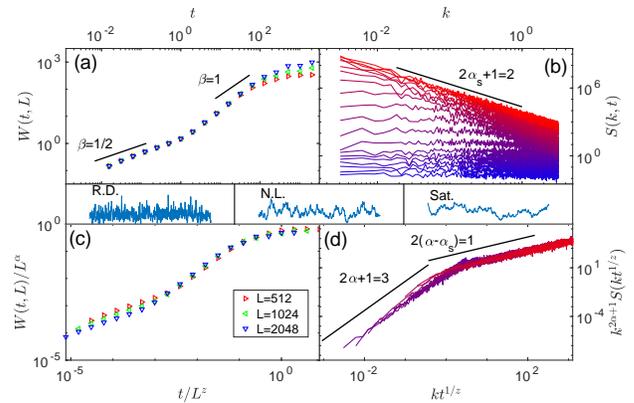}
\caption{Time evolution for (a) the roughness $W(t)$ and (b) structure factor $S(k,t)$ from numerical simulations of Eq.\ \eqref{eq:invkpz}, for $L$ as in the legend. The number of realizations of the noise is $32$, $16$, and $8$ for $L=512,1024,2048$, respectively. Error bars for $W(t)$ \cite{Barras} are smaller than the symbol size. Time values increase bottom to top in (b) and coincide with those used in (a). Data collapses of results for (c) $W(t)$ and (d) $S(k,t)$ obtained using $\alpha=1$, $z=1$, and $\alpha_s=1/2$. Solid lines represent power-law behavior with the indicated values of the exponents. Sample morphologies $h(x,t)$ appear in the middle panels for $L=512$ and times in the random deposition (RD), nonlinear growth (NL), and saturation (Sat.) regimes, left to right. }
\label{fig:WSkpz}
\end{center}
\end{figure}

We proceed with the numerical simulation of Eq.\ \eqref{eq:invkpz}. As the equation depends on two parameters ($\lambda$ and $D$) and three independent rescalings can be done for $x$, $t$, and $h$, {we choose} units so that $\lambda=2$ and {$D=1/2$} without loss of generality. Numerical simulations are carried out using the multistep, predictor-corrector pseudospectral scheme proposed in Ref.\ \cite{Gallego11} and uniform-distributed noise of unit variance \cite{Nota}. We consider a flat initial condition $h(x,t=0)=0$ on a 1D substrate of lateral size $L$, with periodic boundary conditions. Results of our simulations {are shown in Fig.\ \ref{fig:WSkpz}.}
In the middle panels {of the figure}, representative front morphologies $h(x,t)$ are shown for increasing times, left to right. After an initial (random deposition, RD) transient in which height values are uncorrelated in space, the surface enters a time regime dominated by the nonlinear (NL) term, to eventually saturate to steady state at long enough times. More quantitatively, Fig.\ \ref{fig:WSkpz}(a) shows the time evolution of the surface roughness, {namely, the standard deviation of the height field $h(x,t)$,
\begin{equation}
    W(t)= \Big\langle \sqrt{ \frac{1}{L} \int_0^L [h(x,t)-\bar{h}(t)]^2 dx } \Big\rangle,
\label{eq:W}
\end{equation}
where bar denotes spatial average and brackets denote average over different realizations of the noise}, for several values of the system size $L$.
Under kinetic roughening conditions, $W$ grows with time as $W \sim t^{\beta}$ up to a saturation value $W_{\rm sat} \sim L^{\alpha}$ that remains time-independent beyond a time $t_{\rm sat} \sim L^z$, where $\alpha$ is the roughness exponent, $z$ is the dynamic exponent mentioned above, and $\beta=\alpha/z$ \cite{Barabasi95,Krug97}. As seen in the figure, $\beta\simeq 1/2$ for short times, the well-known RD behavior \cite{Barabasi95}. However, for longer times before saturation $\beta\simeq 1$. This behavior is confirmed by the collapse of the $W(t)$ data in Fig.\ \ref{fig:WSkpz}(c) for large $t/L^z$, obtained for different values of $L$ and using $\alpha=z=1$, implying $\beta=1$ \cite{Barabasi95,Krug97}. Note, these exponent values satisfy the so-called Galilean scaling relation, $\alpha+z=2$, associated with the KPZ nonlinearity \cite{Kardar86}, but with exponents very far from the standard 1D KPZ values; in particular, recall that $\alpha_{\rm KPZ}=1/2$ as for the random walk \cite{Barabasi95,Krug97}.

To address (two-point) correlation functions, we consider the height structure factor 
{
\begin{equation}
    S(k,t)=\langle |\tilde{h}(k,t)|^2\rangle ,
\label{eq:PSD}
\end{equation}
}
where tilde denotes space Fourier transform and $k$ is the wave number. While mathematically carrying the same information as real-space correlation functions like the height covariance \cite{Krug97}, this function is particularly informative in the presence of crossover and/or anomalous scaling behavior \cite{Siegert96,Lopez97,Ramasco00}. Results are shown in Fig.\ \ref{fig:WSkpz}(b). The $S(k,t)$ curves for small times in the RD regime are $k$-independent (white noise), as expected for uncorrelated RD behavior. For longer times, power-law behavior develops that behaves asymptotically as $S(k)\sim 1/k^{2\alpha_s+1}$, with $\alpha_s=1/2 \neq \alpha$. Here, $\alpha_s$ is the roughness exponent that characterizes the scaling behavior of e.g.\ the height-difference correlation function
{
\begin{equation}
G(r,t) = \langle (h(x_0,t)-h(x_0+r,t))^2\rangle,
\label{eq:G}
\end{equation}
which behaves as} $G(r,t) \sim r^{2\alpha_s}$ at {\em local} scales $r \ll t^{1/z} \ll L$, while $\alpha$ characterizes fluctuations of {\em global} quantities like $W$ \cite{Barabasi95,Krug97}. 
While, interestingly, the value of the local roughness exponent $\alpha_s$ turns out not to depend on the value of $\nu$,
the fact that $\alpha_s\neq\alpha$ contrasts with the standard KPZ behavior for which $\alpha_{s,{\rm KPZ}}=\alpha_{\rm KPZ}$, and is an indication of intrinsic anomalous scaling \cite{Lopez97,Ramasco00,Cuerno04}. Full confirmation is obtained from the data collapse of the structure factor data also shown in Fig.\ \ref{fig:WSkpz}(d), which agrees with the scaling Ansatz of intrinsic anomalous scaling, i.e.\ \cite{Lopez97,Ramasco00,Cuerno04},
\begin{equation}\label{eq:SkU}
    S(k,t) \sim \frac{s(t/L^z)}{k^{2\alpha_s+1}}, \quad
    s(y) \sim \left\{ 
    \begin{array}{ll}
        y^{2\alpha+1},  & y \ll 1 \\
        y^{2(\alpha-\alpha_s)}, & y \gg 1
    \end{array}
    \right. ,
\end{equation}
again for $\alpha=z=1$ and $\alpha_s=1/2$. Note, this scaling Ansatz retrieves the one satisfied by the KPZ equation for $\nu\neq 0$ when $\alpha=\alpha_s$ and for the corresponding exponent values \cite{Lopez97,Ramasco00,Cuerno04}. This result is unexpected in the context of a previous conjecture \cite{Lopez05} that intrinsic anomalous scaling cannot be asymptotic for continuum models like Eqs.\ \eqref{eq:kpz} or \eqref{eq:invkpz}, which feature local interactions and time-dependent noise (in contrast with, e.g., quenched disorder). Such an expectation was based on perturbative arguments. However, the coupling constant which controls the KPZ scaling behavior, $g=\lambda^2 D/\nu^3$ \cite{Kardar86,Barabasi95,Krug97}, is infinite for Eq.\ \eqref{eq:invkpz}, so that the conjecture does not necessarily apply for this equation.

\subsection{Inviscid stochastic Burgers equation}
\label{sec:Burgers}

As an alternative check of our numerical results, we have considered the space derivative of Eq.\ \eqref{eq:invkpz}, namely, the inviscid stochastic Burgers equation \cite{Burgers74} {for $u=\partial_x h$}, i.e.,
\begin{equation}\label{eq:invburgers}
    \partial_t u = {\lambda} u \partial_x u + \partial_x\eta(x,t) .
\end{equation}
Our strategy is to simulate {Eq.\ \eqref{eq:invburgers}} via the pseudospectral scheme proposed in Ref.\ \cite{Gallego11} and, at each time, obtain the solution of Eq.\ \eqref{eq:invkpz} as the space integral 
{
\begin{equation}\label{eq:uyh}
    h(x,t) = \int_0^x u(x',t) dx'.
\end{equation}
}
This approach has been successfully taken earlier \cite{Rodriguez-Fernandez20} to study the detailed relation between the solutions of the stochastic Burgers and the KPZ equations in the growth regime and at steady state, in the $\nu\neq 0$ case. As noted {in Sec.\ \ref{sec:direct} above, here we can similarly} fix $\lambda=1$ and $D=1/2$ without loss of generality. We integrate Eq.\ \eqref{eq:invburgers} numerically using the scheme proposed in Ref.\ \cite{Gallego11}, using periodic boundary conditions for a 1D system of lateral size $L$, a zero initial condition, and uniform noise of unit variance, {and at each value of time compute $h(x,t)$ via Eq.\ \eqref{eq:uyh}. The results are shown in Fig.\ \ref{fig:CompInvKpz} which, being virtually identical to Fig.\ \ref{fig:WSkpz}, is provided in Appendix \ref{app:1} for the interested reader. Thus, this approach reproduces accurately the results of the direct integration of Eq.\ \eqref{eq:invkpz} already depicted in Fig.\ \ref{fig:WSkpz}, but now for the field $h$ that is computed via Eq.\ \eqref{eq:uyh}}. 

\begin{figure}
\begin{center}
\includegraphics[width=0.85\columnwidth]{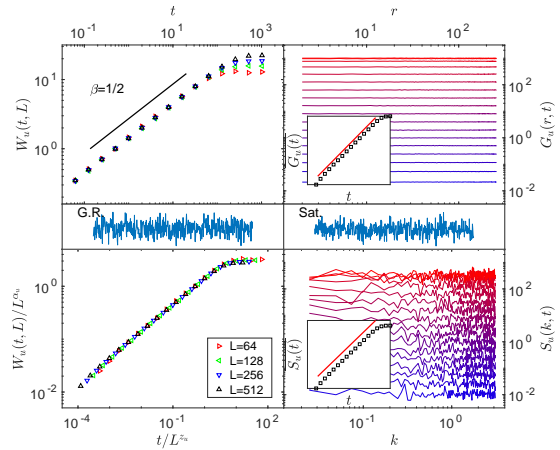}
\caption{\small{Time evolution for the roughness, $W_u(t,L)$ the difference correlation function, $G_u(r,t)$, and the structure factor, $S_u(k,t)$ of the field $u(x,t)$, from numerical simulations of Eq.\ \eqref{eq:invburgers} for values of $L$ as given in the legend of the bottom left panel. The number of realizations of the noise is $64$, $32$, $16$, and $8$ for $L=64,128,256$, and 512, respectively. Error bars for $W_u(t)$ are smaller than the symbol size. The $G_u(r,t)$ and $S_u(k,t)$ data are shown for $L=512$ only. The data collapse of the roughness for $\alpha_u=1/3$ and $z_u=2/3$ is shown in the bottom left panel. Morphologies $u(x)$ for growth regime (G.R.) and saturation (Sat.) are also depicted left to right in the middle panels. Insets of the right panels represent the evolution in time for the averages (denoted by overbars) of $G_u(r,t)$ and $S_u(k,t)$ over $r$ and $k$, respectively, where solid red lines have unit slope.}}
\label{fig:WGSinvburgers}
\end{center}
\end{figure}

{Beyond providing the non-trivial test just discussed for the results of Sec.\ \ref{sec:direct}, the inviscid stochastic Burgers equation displays kinetic roughening behavior which is interesting on its own and we discuss it next.
The results are directly provided by the simulations of Eq.\ \eqref{eq:invburgers} mentioned in the previous paragraph and are shown in Fig.\ \ref{fig:WGSinvburgers}} for the surface roughness $W_u(t)$, surface structure factor $S_u(k,t)$, and difference correlation function $G_u(r,t)$. {Note, in this figure we consider the same formulae, Eqs.\ \eqref{eq:W}, \eqref{eq:PSD}, and \eqref{eq:G}, as employed in Sec.\ \ref{sec:direct}, but now as applied to the field $u(x,t)$, instead of $h(x,t)$.} As seen in Fig.\ \ref{fig:WGSinvburgers}, the behavior for the roughness is notably simple: already starting with the shortest times, $W_u$ increases as $W_u(t) \sim t^{\beta_u}$, to only saturate at sufficiently long times. As borne out from the data collapse also shown in Fig.\ \ref{fig:WGSinvburgers}, the saturated roughness $W_{u,{\rm sat}} \sim L^{\alpha_u}$ and saturation time $t_{u,{\rm sat}}\sim L^{z_u}$ scale as expected for kinetic roughening systems \cite{Barabasi95}. The best collapses for $W_u(t)$ are obtained using scaling exponent values $\alpha_{u} = 1/3$, $\beta_u = 1/2$, and $z_u = 2/3$. Remarkably, {as noted above} $\beta=1/2$ is the value of the growth exponent for the random deposition (RD) process, which produces uncorrelated surfaces and does not saturate to steady state \cite{Barabasi95}. At variance with RD, Eq.\ \eqref{eq:invburgers} does saturate to steady state while similarly producing uncorrelated interfaces. This is clearly seen in the behavior of the difference correlation function and the structure factor, shown in Fig.\ \ref{fig:WGSinvburgers}. While the correlation function is $r$-independent, the spectrum is $k$-independent as well, as for white noise. Hence, the space average $\overline{G}_u(t)$ of $G_u(r,t)$ and the wavenumber average $\overline{S}_u(t)$ of $S_u(k,t)$ are both expected to scale as $W_u^2 \sim t^{2\beta_u}$, as seen in the corresponding insets of Figure \ref{fig:WGSinvburgers}.

\begin{figure}[!t]
\begin{center}
\includegraphics[width=0.9\columnwidth]{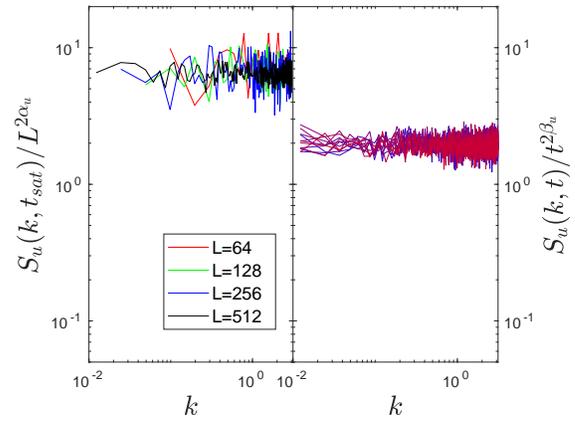}
\caption{\small{Collapses of the structure factor $S_u(k,t)$ data shown in Fig.\ \ref{fig:WGSinvburgers} for the field $u(x,t)$ in Eq.\ \eqref{eq:invburgers}, with respect to the system size $L$ at saturation using $\alpha_u=1/3$ (left panel), and with respect to time $t$ using $\beta_u=1/2$ prior to saturation, with blue to red colors corresponding to increasing times (right panel).}}
\label{fig:coll_WGSukdv}
\end{center}
\end{figure}

The scaling behavior of the structure factor for steady state solutions of $u$ in Eq.\ \eqref{eq:invburgers} in consistent with the form
\begin{equation}\label{eq:SkUMig}
    S_u(k,L) \sim \frac{L^{2\alpha_u}}{k^{2\alpha_{s,u}+d}},
\end{equation}
as seen in the left panel of Fig.\ \ref{fig:coll_WGSukdv} for $\alpha_u=1/3$, $\alpha_{s,u}=-1/2$, and $d=1$. This behavior coincides with that described in Refs.\ \cite{Rodriguez14,Rodriguez15} for time series (of length $L$) of stochastic pulses. In our case, we can generalize Eq.\ \eqref{eq:SkUMig} for times prior to saturation as
\begin{equation}\label{eq:SkUNos}
    S_u(k,t) \sim \frac{L^{2\alpha_u}}{ k^{2\alpha_{s,u}+d}} f_S(t/L^z),
\end{equation}
{with}
\begin{equation}
    f_S(y) \sim \left\{ 
    \begin{array}{ll}
        y^{2\beta_u },  & y \ll 1 \\
        \rm{cnst}, & y \gg 1
    \end{array}
    \right. .
\end{equation}
Indeed, the scaling of $S_u$ with $t$ predicted by Eq.\ \eqref{eq:SkUNos} is verified in the right panel of Fig.\ \ref{fig:coll_WGSukdv}.

While {$\alpha_u$ and $z_u$} satisfy the Galilean scaling relation $\alpha+z=1$ expected for the stochastic Burgers equation \cite{Rodriguez-Fernandez20}, they conspicuously differ from the values obtained for the viscous case of this equation, which are $\alpha_{\rm visc}=-1/2$ and $z_{\rm visc}=3/2$ \cite{Rodriguez-Fernandez20}. {Moreover}, the exponent relation expected between Eqs.\ \eqref{eq:invkpz} and \eqref{eq:invburgers} \cite{Rodriguez-Fernandez20} holds for {the local roughness exponent as $\alpha_{s,u}=\alpha_s-1$, but not for the global roughness exponent, since $\alpha_u \neq \alpha-1$}. It is also remarkable that $S_u(k,t)$ and $G_u(r,t)$ are like those of white noise for all $t$, making $u(x,t)$ a random deposition-like process in which saturation occurs. Its steady state is like that obtained in Refs.\ \cite{Rodriguez14,Rodriguez15} for a stochastic model of independent pulses. Overall, this nontrivial dynamics for the slope field $u$ induces the anomalous scaling observed for $h$, as argued for on general grounds for intrinsically anomalous kinetic roughening systems \cite{Schroeder93,DasSarma94,Lopez99}.


\section{Fluctuation statistics}\label{sec:fluctuation}

\begin{figure}[!t]
\begin{center}
\includegraphics[width=1.00\columnwidth]{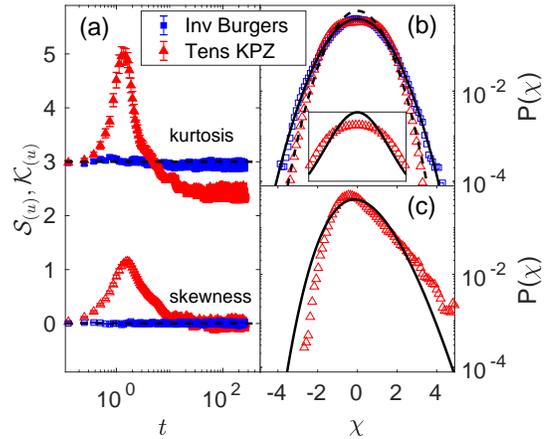}
\caption{(a) Time {evolution} for the fluctuation skewness ${\cal S}$ (empty) and kurtosis ${\cal K}$ (filled), and (b-c) PDF of normalized field fluctuations [$\chi = (\phi-\bar\phi)/{\rm std}(\phi)$, where bar is space average] of Eq.\ \eqref{eq:invburgers} [$\phi=u$, square symbols in all panels] and Eq.\ \eqref{eq:invkpz} [$\phi=h$, triangle symbols in all panels], for $L=512$ and 100 noise realizations. For error bars, see \cite{Barras}. (b) PDF at steady state. Inset: linear zoom of the central part of the exact Gaussian and the stationary PDF for tensionless KPZ. (c) PDF for the tensionless KPZ fluctuations at the maximum of ${\cal S}$ and ${\cal K}$. Solid lines correspond to the exact Gaussian (b) and GOE-TW (c) distributions. The dashed line in (b) corresponds to the large-$\chi$ fit $P(\chi) = 0.6 \exp(-0.7\chi^2)$. }
\label{fig:pdf}
\end{center}
\end{figure}

The most recent developments in the context of 1D KPZ universality (see e.g.\ Refs.\  \cite{Halpin-Healy15,Takeuchi18,Rodriguez-Fernandez19,Rodriguez-Fernandez20,Rodriguez-Fernandez21} and other therein) underscore the importance of characterizing the PDF of the field fluctuations to unambiguously determine the kinetic roughening universality class. Figure \ref{fig:pdf} shows our results for Eqs.\ \eqref{eq:invkpz} and \eqref{eq:invburgers}; recall that we are using flat initial conditions in both cases.
The fluctuations for $u(x,t)$ are Gaussian along its full dynamics, as manifested by the values of its skewness
{
\begin{equation}\label{eq:skew}
    {\cal S}_u(t)=\frac{ 1}{W_u^3(t)} \Big\langle \frac{1}{L} \int_0^L [u(x,t)-\bar{u}(t)]^3 dx \Big\rangle
\end{equation}
}
and kurtosis
{
\begin{equation}\label{eq:kurt}
    {\cal K}_u(t)=\frac{ 1}{W_u^4(t)} \Big\langle \frac{1}{L} \int_0^L [u(x,t)-\bar{u}(t)]^4 dx \Big\rangle,
\end{equation}
where $W_u$ is the roughness of the $u(x,t)$ field,} and by the full PDF.
{Note that the exact values for a Gaussian distribution are $\mathcal{S}_{\rm Gauss}=0$ and $\mathcal{K}_{\rm Gauss}=3$. The Gaussian behavior that we obtain {in Fig.\ \ref{fig:pdf}} for the fluctuations of $u$} is much like the behavior obtained for the Burgers equation with $\nu\neq 0$ subject either to non-conserved \cite{Rodriguez-Fernandez19} or to conserved \cite{Rodriguez-Fernandez20} noise. 

For the tensionless KPZ equation, Eq.\ \eqref{eq:invkpz}, starting out from Gaussian values at very short times, both ${\cal S}(t)$ and ${\cal K}(t)$ { [defined as in Eqs.\ \eqref{eq:skew} and \eqref{eq:kurt}, but for $h$ instead of $u$]} increase sharply reaching finite maxima; the PDF at this time departs clearly both from a Gaussian and from the Tracy-Widom distribution of the largest eigenvalue of random matrices in the Gaussian Orthogonal Ensamble (GOE-TW). The latter PDF is expected in the growth regime of the $\nu\neq0$ KPZ equation for our present boundary conditions \cite{Kriechebauer10,Corwin12,Halpin-Healy15,Takeuchi18}. The stationary state fluctuations for Eq.\ \eqref{eq:invkpz} become symmetric (zero skewness) but their kurtosis does not take the Gaussian value expected for $\nu\neq0$ \cite{Barabasi95,Krug97} because the PDF is flatter than a Gaussian in its central part. Nonetheless, the tails agree well with Gaussian decay, as suggested by the fit provided in Fig.\ \ref{fig:pdf}(b), so $\mathcal{K}\to 0$ for $L\to\infty$ could be expected, as in the $\nu\neq0$ Burgers equation \cite{Rodriguez-Fernandez19,Rodriguez-Fernandez20}. For the nonlinear $\nu\neq0$ KPZ equation in 1D, the stationary PDF is known to be the same as that of the linear ($\lambda=0$) case as a consequence of a fluctuation-dissipation theorem \cite{Barabasi95,Krug97,Cartes22}. {For the interested reader and as a further cross-check, Appendix \ref{app:1} provides the full analysis of the statistics of the $h$ fluctuations evaluated via Eq.\ \eqref{eq:uyh}. The results are shown in Fig.\ \ref{fig:pdf_sm}, which is virtually identical to Fig.\ \ref{fig:pdf}.}

\section{Universality classes }\label{sec:universality}

The universality class that ensues for the tensionless KPZ equation has been observed earlier without reference to Eq.\ \eqref{eq:invkpz} itself. The same set of (Galilean-invariant, ballistic) global exponent values $\alpha=z=1$ has been obtained for a non-local generalization of the KPZ equation \cite{Nicoli09} that quantitatively describes experimental thin films grown by chemical vapor deposition \cite{Nicoli12}. However, in this case $\alpha_s=\alpha=1$ so that a regular dynamic scaling Ansatz is fulfilled, in contrast with our present result {of intrinsic anomalous scaling} for Eq.\ \eqref{eq:invkpz}. Better agreement is found in simulations of a 2D discrete growth model associated with an invasion percolation process, under conditions that suppress fluid trapping \cite{Asikainen02,Asikainen02b}. The kinetic roughening of the corresponding height field is intrinsically anomalous, with exponents $\alpha \simeq 0.99$, $\beta\simeq 1$, $z\simeq 0.99$, and $\alpha_s\simeq 0.51$ \cite{Asikainen02,Asikainen02b}, very similar to the values we obtain for Eq.\ \eqref{eq:invkpz} at long times. {Additional members of the universality classes of the tensionless KPZ equation and of the stochastic inviscid Burgers equation are discussed next.}

\subsection{Stochastic KdV equation }\label{sec:kdv}

Further, the same full exponent set and scaling Ansatz {that we find for Eq.\ \eqref{eq:invkpz} or for Eq.\ \eqref{eq:invburgers}} can also be found for other relevant continuum models. 
\begin{figure}[!t]
\begin{center}
\includegraphics[width=1.00\columnwidth]{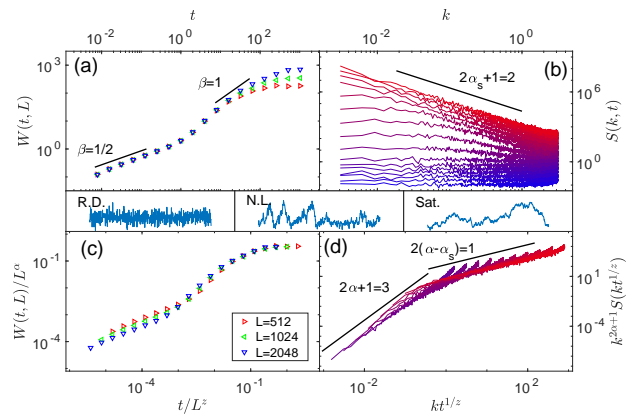}
\caption{Time evolution for the roughness $W(t)$ and structure factor $S(k,t)$ from {direct} numerical simulations of Eq.\ \eqref{eq:kdvkpz}, for $L$ as given in the bottom left panel legend and $c=1$. The description of the various panels and the values of the scaling exponents are as in Fig.\ \ref{fig:WSkpz}.}
\label{fig:WSkdvkpz}
\end{center}
\end{figure}
Specifically, consider
\begin{equation}\label{eq:kdvkpz}
    \partial_t h = c \partial_x^3 h + \frac{1}{2}(\partial_x h)^2 + \eta(x,t) ,
\end{equation}
where $c$ is a parameter and $\eta$ is as in Eq.\ \eqref{eq:kpz}, whose space derivative is the stochastic Korteweg-de Vries (KdV) equation
\begin{equation}\label{eq:kdv}
    \partial_t u = c \partial_x^3 u + u \partial_x u + \partial_x\eta(x,t) .
\end{equation}
The deterministic KdV equation is a paradigmatic model of weakly-nonlinear waves and is well-known to be exactly solvable \cite{Whitham99}. Equation \eqref{eq:kdv} generalizes it by adding conserved, time-dependent noise, see  e.g.\ Refs.\ \cite{Debussche06,DaPrato14} for related systems. 

{Completely analogous to the approach we employed at the beginning of Sec.\ \ref{sec:Burgers}, numerical simulations of Eq.\ \eqref{eq:kdv} can be employed to study the behavior of its space integral, Eq.\ \eqref{eq:kdvkpz}. Results are presented in Fig.\ \ref{fig:WSkdvkpz}, and indicate} a scaling behavior which is quite close to that obtained for the tensionless KPZ equation, Eq.\ \eqref{eq:invkpz}. The time regimes and the values of the scaling exponents are exactly like those seen in Fig.\ \ref{fig:WSkpz}, which confirms the intuitive expectation that the dispersive linear term with parameter $c$ appreciably influences neither the value of $S(k,t)$, nor that of $W^2(t)=\int S(k,t) dk$ for any value of time.

{On the other hand, the results for the dynamics of the $u$ field from our numerical simulations of Eq.\ \eqref{eq:kdv} are shown in Fig.\ \ref{fig:WGSukdv} which, being quantitatively very similar to Fig.\ \ref{fig:WGSinvburgers}, is shown in Appendix \ref{app:2}. The kinetic roughening behavior obtained is identical to that discussed in Sec.\ \ref{sec:Burgers} for Eq.\ \eqref{eq:invburgers}. Hence, 
the stochastic KdV equation, Eq.\ \eqref{eq:kdv}, is in the universality class
of the inviscid stochastic Burgers equation.}

\section{Conclusions and outlook }\label{sec:conclusions}

Summarizing, we have elucidated a well-defined universality class for the tensionless KPZ equation for one-dimensional interfaces that encompasses additional discrete and continuum models. The former include at least systems related to invasion percolation \cite{Asikainen02,Asikainen02b}, while the latter include models related with the KdV equation, Eq.\ \eqref{eq:kdvkpz}, with time-dependent noise. {Along this process, we have additionally elucidated the universality class of the related inviscid stochastic Burgers equation.}

Despite Eq.\ \eqref{eq:invkpz} featuring short-range interactions and standard time-dependent noise, its scaling behavior turns out to be intrinsically anomalous, with different roughness exponents controlling height fluctuations at local and global length scales, and with ballistic propagation of correlations featuring a dynamic exponent $z=z_{\rm b}=1$ (as also recently reported in Ref.\ \cite{Cartes22}) which can be a test for conformal invariance \cite{Henkel10} in suitable parameter regions for e.g.\ quantum spin chains \cite{Gopalakrishnan19,Ljubotina19,DeNardis21,Wei21}. 
As noted above, a recent experiment has measured both the $z=z_{\rm b}=1$ value and $z=z_{\rm KPZ}=3/2$ (as for the viscous KPZ equation) in different parameter regions for the evolution of a quantum spin chain \cite{Wei21}. In view of our present results, these observations might be understood within a single framework, corresponding to $\nu$ taking zero or non-zero values, respectively, in the effective KPZ equation relevant for different experimental conditions. Confirmation could be achieved via measurement of e.g.\ additional scaling exponents, like $\alpha$ and/or $\alpha_s$, in parameter regions corresponding to the observed ballistic behavior, and of the accurate time evolution of the skewness and kurtosis of the fluctuations.

The dynamical behavior obtained for Eq.\ \eqref{eq:invkpz} at large scales is obviously induced by the nonlinear term, and may be intuitively rationalized, bearing in mind that (in the deterministic limit) the latter propagates an interface with a constant speed along the local normal direction as in the eikonal equation \cite{Sethian05}, implementing a Huygens principle, while no other competing (deterministic) relaxation term exists in Eq.\ \eqref{eq:invkpz}, in contrast with  the viscous KPZ equation, Eq.\ \eqref{eq:kpz}. Notably, a tensionless equation like Eq.\ \eqref{eq:invkpz}, but defined on a medium with suitable quenched disorder, leads to standard viscous KPZ behavior \cite{Santalla15,Santalla17}, including Tracy-Widom statistics which reflect global system constraints \cite{Kriechebauer10,Corwin12,Halpin-Healy15,Takeuchi18}. Possibly, the neglect of self-intersections in the interface evolution \cite{Santalla15,Santalla17} may be inducing an effective non-zero value of the surface tension. 

From an analytical point of view, as noted above Eq.\ \eqref{eq:invkpz} defies the exact solutions obtained in the KPZ case; conversely, the existence of such solutions guarantees that, no matter how small $\nu \neq 0$ is, standard KPZ asymptotics ensues, as the Cole-Hopf transformation can then be applied. The $\nu=0$ condition also challenges perturbative dynamical normalization group approaches \cite{Kardar86,Tauber14,Golubovic91,Cuerno95}, as its fixed point corresponds to an infinite coupling constant $g$. Improved non-perturbative approaches seem required to access the asymptotic behavior analytically and confirm the non-renormalization of $\nu$ for a bare $\nu=0$ which seems implied by our numerical results.


A natural issue concerns the extension of our present results to higher dimensions. Note that, when $x \in \mathbb{R}^d$ with $d>1$, if the relevant $h(x,t)$ field is to remain a {\em scalar}, the straightforward 1D relation is lost to an analogous scalar $u(x,t)$ field which satisfies some Burgers-like equation. Although scalar analogs of the Burgers equation are possible in higher dimensions \cite{Vivo14,Rodriguez-Fernandez19}, we have not been successful in following this route to efficiently integrate the corresponding generalization of Eq.\ \eqref{eq:invkpz}, while preliminary direct simulations \cite{Rodriguez-Fernandez22,Rodriguez-Fernandez22c} for $d=2$ suggest finite-time blow-up behavior akin to previous observations \cite{Tabei04,Bahraminasab04}. And although the KdV equation itself conspicuously exemplifies the difficulty of extending its fascinating 1D behavior \cite{Whitham99} to higher dimensions, we consider this as a relevant open issue for Eq.\ \eqref{eq:invkpz}.


\begin{acknowledgments}
This work has been partially supported by Ministerio de Ciencia, Innovaci\'on y Universidades (Spain), Agencia Estatal de Investigaci\'on (AEI, Spain), and Fondo Europeo de Desarrollo Regional (FEDER, EU) through Grants No.\ PGC2018-094763-B-I00 and No.\ PID2019-106339GB-I00, and by Comunidad de Madrid (Spain) under the Multiannual Agreements with UC3M in the line of Excellence of University Professors (EPUC3M14 and EPUC3M23), in the context of the V Plan Regional de Investigaci\'on Cient\'{\i}fica e Innovaci\'on Tecnol\'ogica (PRICIT). E.\ R.-F.\ acknowledges financial support through contract No.\ 2022/018 under the EPUC3M23 line.
\end{acknowledgments}

\appendix


\section{Tensionless KPZ equation as the integral field of the inviscid Burgers equation}\label{app:1}

\begin{figure}[!t]
\begin{center}
\includegraphics[width=1.00\columnwidth]{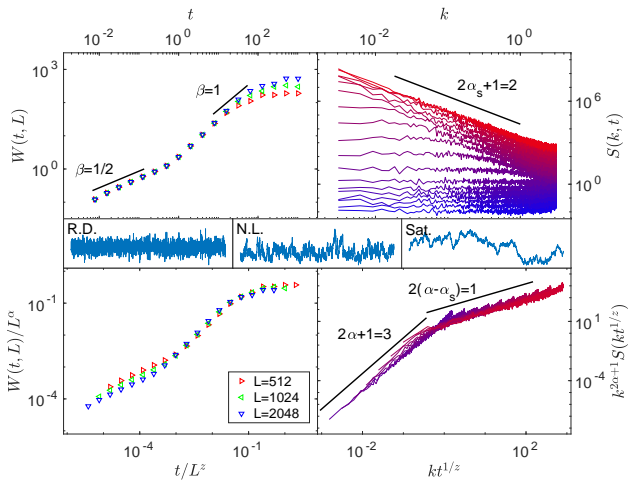}
\caption{\small{Top panels: Time evolution for the roughness $W(t)$ (left) and structure factor $S(k,t)$ (right) of the field $h(x,t)=\int_0^x u(x,t) \ dx$, from numerical simulations of Eq.\ \eqref{eq:invburgers}, for $L$ as in the legend (bottom left panel). The number of realizations of the noise is $32$, $16$, and $8$ for $L=512,1024$, and 2048, respectively. Error bars for $W(t)$ are smaller than the symbol size. Time values increase bottom to top in the right panel and coincide with those used in the left panel. Bottom panels: Data collapses of results for $W(t)$ (left) and $S(k,t)$ (right) obtained using $\alpha=1$, $z=1$, and $\alpha_s=1/2$. Solid lines represent power-law behavior with the indicated values of the exponents. Sample morphologies $h(x,t)$ appear in the middle panels for times in the random deposition (RD), nonlinear growth (NL), and saturation (Sat.) regimes, left to right. }}
\label{fig:CompInvKpz}
\end{center}
\end{figure}

\begin{figure}[!t]
\begin{center}
\includegraphics[width=1.00\columnwidth]{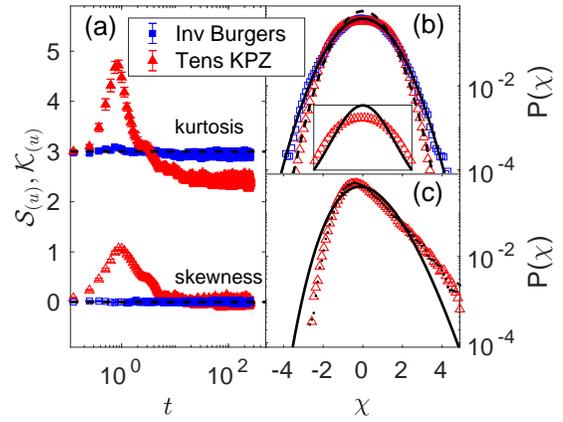}
\caption{(a) Time evolution of the fluctuation skewness ${\cal S}$ (empty) and kurtosis ${\cal K}$ (filled), and (b-c) PDF of normalized field fluctuations [$\chi = (\phi-\bar\phi)/{\rm std}(\phi)$, where bar is space average] of $\phi=u$ (square symbols in all panels) and $\phi=h=\int_0^x u(x') dx'$ (triangle symbols in all panels) from numerical simulations of Eq.\ \eqref{eq:invburgers} for $L=512$ and 100 noise realizations. (b) PDF at steady state. Inset: linear zoom of the central part of the exact Gaussian and the stationary PDF for tensionless KPZ. (c) PDF for the tensionless KPZ fluctuations at the maximum of ${\cal S}$ and ${\cal K}$. Solid lines correspond to the exact Gaussian (b) and GOE-TW (c) distributions. The dashed line in (b) corresponds to the large-$\chi$ fit $P(\chi) = 0.6 \exp(-0.7\chi^2)$. The dotted line in (c) shows the results obtained by direct integration of the tensionless KPZ equation, see Fig.\ \ref{fig:pdf}. }
\label{fig:pdf_sm}
\end{center}
\end{figure}

The numerical simulations performed in {Sec.\ \ref{sec:Burgers}} for the inviscid Burgers equation, Eq.\ \eqref{eq:invburgers}, also provide us with a consistency check of the results obtained in Sec.\ \ref{sec:direct} for the direct integration of the tensionless KPZ equation. Specifically, our strategy is to simulate the $u$ field of the inviscid Burgers equation, Eq.\ \eqref{eq:invburgers} and, at each time, obtain the solution of the tensionless KPZ equation, Eq.\ \eqref{eq:invkpz}, as the space integral $h(x,t) = \int_0^x u(x',t) dx'$. A similar strategy was successfully employed in Ref.\ \cite{Rodriguez-Fernandez20} to study the relation between the viscous ($\nu\neq0$) Burgers and KPZ equations. Results are shown in Figs.\ \ref{fig:CompInvKpz} and \ref{fig:pdf_sm} {which, being virtually identical to, lead to the same conclusions as those obtained from, Figs.\ \ref{fig:WSkpz} and \ref{fig:pdf}, respectively}.

\section{KdV equation with Burgers nonlinearity and conserved noise}\label{app:2}

\begin{figure}[!t]
\begin{center}
\includegraphics[width=0.8\columnwidth]{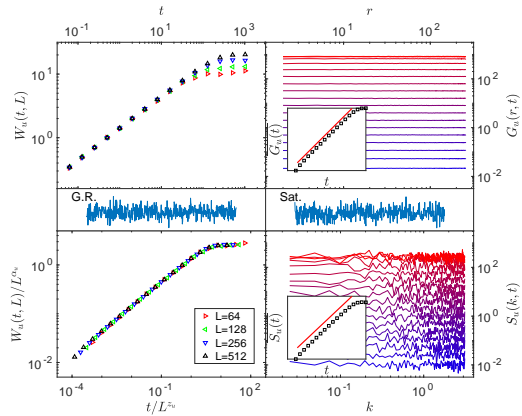}
\caption{\small{Time evolution for the roughness, $W_u(t,L)$, the difference correlation function, $G_u(r,t)$, and the structure factor, $S_u(k,t)$ of the field $u(x,t)$, from numerical simulations of Eq.\ \eqref{eq:kdv} for $c=1$ and values of $L$ as given in the legend of the bottom left panel. The number of realizations of the noise is $64$, $32$, $16$, and $8$ for $L=64,128,256$, and 512, respectively. Error bars for $W(t)$ are smaller than the symbol size. The $G_u(r,t)$ and $S_u(k,t)$ data are shown for $L=512$ only. The data collapse of the roughness for $\alpha=1/3$ and $z=2/3$ is shown in the bottom left panel. Morphologies $u(x)$ for growth regime (G.R.) and saturation (Sat.) are also depicted left to right in the middle panels. Insets of the right panels represent the evolution in time for the averages (denoted by overbars) of $G_u(r,t)$ and $S_u(k,t)$ over $r$ and $k$, respectively where solid red lines have unit slope.}}
\label{fig:WGSukdv}
\end{center}
\end{figure}

We consider the numerical simulation of the KdV equation with Burgers nonlinearity and conserved noise,
Eq.\ \eqref{eq:kdv}. We use the same numerical method and conditions as we have used for the integration of Eq.\ \eqref{eq:invburgers} in Sec.\ \ref{sec:Burgers}. Results are shown in Fig.\ \ref{fig:WGSukdv} and lead to the same conclusions as those obtained from Fig.\ \ref{fig:WGSinvburgers} for the inviscid Burgers equation, Eq.\ \eqref{eq:invburgers}. We conclude that the universality class of Eq.\ \eqref{eq:kdv} is the same as that of the inviscid Burgers equation.

\end{document}